\documentclass [a4paper,twoside,11pt]{article}
\usepackage{a4wide,amsmath,amssymb,latexsym}
\usepackage{delarray,graphics,graphpap,alltt,graphicx}
\usepackage[latin1]{inputenc}       
\usepackage{algorithm}       
\usepackage{algorithmic}     
\usepackage {a4wide,amsmath,amssymb,latexsym,graphicx}
\usepackage{rotating}

\newcommand{\roC}{\rho _C}

\newtheorem{theorem}{Theorem}[section]

\newtheorem{lemma}[theorem]{Lemma}
\newtheorem{corollary}[theorem]{Corollary}
\newtheorem{claim}[theorem]{Claim}
\newtheorem{remark}[theorem]{Remark}

\newenvironment{definition}
{ {\noindent {\bf Definition.}} } {  }

\newenvironment{proposition*}[1]
{ {\noindent {\bf Proposition #1.}} } {  }

\newcommand{\bool}[1]{\{ 0,1\}^{#1}}

\newcommand{\N}{\mathbb{N}}

\def\squareforqed{\hbox{\rlap{$\sqcap$}$\sqcup$}}
\def\qed{\ifmmode\squareforqed\else{\unskip\nobreak\hfil
\penalty50\hskip1em\null\nobreak\hfil\squareforqed
\parfillskip=0pt\finalhyphendemerits=0\endgraf}\fi}

\title{Pushdown Compression}
\date{}

\author{Pilar Albert \footnote{Dept. de Informática
 e Ingeniería de Sistemas
, Universidad de Zaragoza. Edificio Ada Byron, María de Luna 1 -
E-50018 Zaragoza (Spain). Email: \{mpalbert, elvira\}(at)unizar.es.
Research supported in part by Spanish Government MEC
   Project TIN 2005-08832-C03-02 and by Aragón Government Dept. Ciencia, Tecnología y Universidad,
   subvención destinada a la formación de personal
   investigador-B068/2006.}, Elvira Mayordomo \footnotemark[\value{footnote}], Philippe Moser, \footnote{Dept. of Computer Science, National University of
Ireland, Maynooth. Maynooth, Co. Kildare. Ireland. Email:
pmoser(at)cs.nuim.ie} and Sylvain Perifel \footnote{LIP, École
Normale Supérieure de Lyon. UMR 5668 ENS Lyon, CNRS, UCBL, INRIA.
Email: sylvain.perifel(at)ens-lyon.fr.} }
\date{}

\begin{document}
\maketitle


\begin{abstract}

The pressing need for efficient compression schemes for XML
documents has recently been focused on stack computation
\cite{HarSha06, LeaEng07}, and in particular calls for a formulation
of information-lossless stack or pushdown compressors that allows a
formal analysis of their performance and a more ambitious use of the
stack in XML compression, where so far it is mainly connected to
parsing mechanisms. In this paper we introduce the model of pushdown
compressor, based on pushdown transducers that compute a single
injective function while keeping the widest generality regarding
stack computation.


The celebrated Lempel-Ziv algorithm  LZ78 \cite{LemZiv78} was
introduced as a general purpose compression algorithm that
outperforms finite-state compressors on all sequences. We compare
the performance of the Lempel-Ziv algorithm with that of the
pushdown compressors, or compression algorithms that can be
implemented with a pushdown transducer. This comparison is made
without any a priori assumption on the data's source and considering
the asymptotic compression ratio for infinite sequences. We prove
that Lempel-Ziv is incomparable with pushdown compressors.
\end{abstract}



\section*{Keywords}
Finite-state compression, Lempel-Ziv algorithm, pumping-lemma,
pushdown compression, XML document.


\section{Introduction}

The celebrated result of Lempel and Ziv \cite{LemZiv78} that their
algorithm is asymptotically better than any finite-state compressor
is one of the major theoretical justifications of this widely used
algorithm. However, until recently the natural extension of
finite-state to pushdown compressors has received much less
attention.

XML is rapidly becoming a standard for the creation and parsing of
documents, however, a significant disadvantage is document size,
even more since present day XML databases are massive. Since 1999
the design of new compression schemes for XML is an active area
where the use of syntax directed compression is specially adequate,
that is, compression performed with stack memory \cite{HarSha06,
LeaEng07}.


On the other hand the work done on stack transducers has been basic
and very connected to parsing mechanisms. Transducers were initially
considered by Ginsburg and Rose in \cite{GinRos66} for language
generation, further corrected in \cite{GinRos68}, and summarized in
\cite{AuBeBo07}. For these models the role of nondeterminism is
specially useful in the concept of $\lambda$-rule, that is a
transition in which a symbol is popped from the stack without
reading any input symbol.

In this paper we introduce the concept of pushdown compressor as the
most general stack transducer that is compatible with
information-lossless compression. We allow the full power of
$\lambda$-rules while having a deterministic (unambiguous) model.
The existence of endmarkers is discussed, since it allows the
compressor to move away from mere prefix extension by exploiting
$\lambda$-rules.


The widely-used Lempel-Ziv algorithm LZ78 \cite{LemZiv78} was
introduced as a general purpose compression algorithm that
outperforms finite-sate compressors on all sequences when
considering the asymptotic compression ratio. This means that for
infinite sequences, the algorithm attains the (a posteriori) finite
state or block entropy. If we consider an ergodic source, the
Lempel-Ziv compression coincides exactly with the entropy of the
source with high probability on finite inputs. This second result is
useful when the data source is known, whereas it is not very
informative for general inputs, specially for the case of infinite
sequences (notice that an infinite sequence is Lempel-Ziv
incompressible with probability one). For the comparison of
compression algorithms on general sequences, either an experimental
or a formal approach is needed, such as that used in
\cite{LatStr97}. In this paper we follow \cite{LatStr97} using a
worse case approach, that is, we consider asymptotic performance on
every infinite sequence.

We compare the performance of the Lempel-Ziv algorithm with that of
the pushdown-compressors, or compression algorithms that can be
implemented with a pushdown transducer. This comparison is made
without any a priori assumption on the data's source and considering
the asymptotic  compression ratio for infinite sequences.

We prove that Lempel-Ziv compresses optimally a sequence that no
pushdown transducer compresses at all, that is, the Lempel-Ziv and
pushdown compression ratios of this sequence are $0$ and $1$,
respectively. For this result, we develop a powerful nontrivial
pumping-lemma, that has independent interest since it deals with
families of pushdown transducers, while known pumping-lemmas are
restricted to recognizing devices \cite{AuBeBo07}.

In fact, Lempel-Ziv and pushdown compressing algorithms are
incomparable, since we construct a sequence that is very close to
being Lempel-Ziv incompressible while the pushdown compression ratio
is at most one half. While Lempel-Ziv is universal for finite-state
compressors, our theorem implies a strong non-universality result
for Lempel-Ziv and pushdown compressors.

The paper is organized as follows. Section 2 contains some
preliminaries. In section \ref{sec_PDcompression}, we present our
model of pushdown compressor with its basic properties and notation.
In section \ref{sec_LZoutperformsPD} we show that there is a
sequence on which Lempel-Ziv outperforms pushdown compressors and in
section \ref{sec_LZnouniversalPD} we show that Lempel-Ziv and
pushdown compression are incomparable. We finish with a brief
discussion of connections and consequences of these results for
dimension and prediction algorithms.



Our proofs appear in the appendix.


\section{Preliminaries}\label{sec_preliminaries}
We write $\mathbb{Z}$ for the set of all integers, $\mathbb{N}$ for
the set of all nonnegative integers and $\mathbb{Z^+}$ for the set
of all positive integers. Let $\Sigma$ be a finite alphabet, with
$|\Sigma| \geq 2$. $\Sigma^*$ denotes the set of finite strings, and
$\Sigma ^\infty $ the set of infinite sequences. We write $|w|$ for
the length of a string $w$ in $\Sigma ^*$. The empty string is
denoted by $\lambda $. For $S$ $\in $ $\Sigma ^\infty $ and $i,j$
$\in $ $\mathbb{N}$, we write $S[i..j]$ for the string consisting of
the $i^{\textrm{th}}$ through $j^{\textrm{th}}$ bits of $S$, with
the convention that $S[i..j]=\lambda $ if $i>j$, and $S[0]$ is the
leftmost bit of $S$. We write $S[i]$ for $S[i..i]$ (the
$i^{\textrm{th}}$ bit of $S$). For $w$ $\in $ $\Sigma ^*$ and $S$
$\in $ $\Sigma ^\infty $, we write $w\sqsubseteq S$ if $w$ is a
prefix of $S$, i.e., if $w=S[0..|w|-1]$. Unless otherwise specified,
logarithms are taken in base $|\Sigma|$. For a string $x$, $x^{-1}$
denotes $x$ written in reverse order. We use $f(x)=\perp $ to denote
that function $f$ is undefined on $x$.

    Let us give a brief description of the Lempel-Ziv (LZ) algorithm \cite{LemZiv78}. Given an input
$x\in\Sigma^*$, LZ parses $x$ in different phrases $x_i$,
    i.e., $x=x_1 x_2 \ldots x_n$ ($x_i\in\Sigma^*$) such that
    every prefix $y\sqsubset x_i$, appears before $x_i$ in the parsing (i.e. there exists $j<i$ s.t. $x_j=y$).
    Therefore for every $i$, $x_i = x_{l(i)}b_i$ for $l(i)<i$ and
    $b_i\in\Sigma$. We sometimes denote the number of phrases in the
    parsing of $x$ as $P(x)$.

    LZ encodes $x_i$ by a prefix free encoding of $l(i)$ and the
    symbol $b_i$, that is, if $x=x_1 x_2 \ldots x_n$ as before, the
    output of LZ on input $x$ is
    \[LZ(x)= c_{l(1)}b_1 c_{l(2)}b_2 \ldots c_{l(n)}b_n\] where
    $c_i$ is a prefix-free coding of $i$ (and $x_0=\lambda$).

    LZ is usually restricted to the binary alphabet, but the
    description above is valid for any $\Sigma$.

    For a sequence $S\in\Sigma^{\infty}$, the LZ infinitely often compression ratio is given by
    $$
        \rho_{LZ}(S) = \liminf_{n\rightarrow\infty} \frac{|LZ(S[0\ldots n-1])|}{n\log_2(|\Sigma|)}        .
    $$

We also consider the almost everywhere compression ratio
$$R_{LZ}(S) = \limsup_{n\rightarrow\infty} \frac{|LZ(S[0\ldots n-1])|}{n\log_2(|\Sigma|)}.$$


\section{Pushdown compression}\label{sec_PDcompression}
\begin{definition}A {\itshape pushdown compressor} $(PDC)$
is a 7-tuple
\begin{center}
    $C=(Q, \Sigma , \Gamma , \delta , \nu , q_0, z_0)$
\end{center}
where\begin{itemize}
\item $\Sigma$ is the finite input alphabet
       \item $Q$ is a finite set of states
       \item $\Gamma $ is the finite stack alphabet
       \item $\delta :Q\times (\Sigma \cup \{\lambda \})\times \Gamma \rightarrow Q\times \Gamma
       ^*$ is the transition function
       \item $\nu :Q\times (\Sigma \cup \{\lambda \})\times \Gamma \rightarrow
       \Sigma ^*$ is the output function
       \item $q_0$ $\in $ $Q$ is the initial state
       \item $z_0$ $\in $ $\Gamma $ is the start stack symbol
     \end{itemize}
\end{definition}
We write $\delta =(\delta _Q, \delta _{\Gamma ^*})$. Note that the
transition function $\delta $ accepts $\lambda $ as an input
character in addition to elements of $\Sigma$, which means that $C$
has the option of not reading an input character while altering the
stack. In this case $\delta (q, \lambda, a)=(q', \lambda )$, that
is, we pop the top symbol of the stack. To enforce determinism, we
require that at least one of the following hold for all $q$ $\in $
$Q$ and $a$ $\in $ $\Gamma $:\begin{itemize}
            \item $\delta (q,\lambda ,a)=\perp $
            \item $\delta (q,b,a)=\perp $ for all $b$ $\in $ $\Sigma$
            \end{itemize}
We restrict $\delta $ so that $z_0$ cannot be removed from the stack
bottom, that is, for every $q$ $\in $ $Q$, $b$ $\in $ $\Sigma \cup
\{\lambda \}$, either $\delta (q,b,z_0)=\perp $, or $\delta
(q,b,z_0)=(q',vz_0)$, where $q'$ $\in $ $Q$ and $v$ $\in $ $\Gamma
^*$.

There are several natural variants for the model of pushdown
transducer \cite{AuBeBo07}, both allowing different degrees of
nondeterminism and computing partial (multi)functions by requiring
final state or empty stack termination conditions. Our purpose is to
compute a total and well-defined (single valued) function in order
to consider general-purpose, information-lossless compressors.

Notice that we have not required here or in what follows that the
computation should be invertible by another pushdown transducer,
which is a natural requirement for practical compression schemes.
Nevertheless the unambiguity condition of a single computation per
input gives as a natural upper bound on invertibility.

We use the extended transition function $\delta ^{*}:Q\times
\Sigma^* \times \Gamma ^+\rightarrow Q\times \Gamma
    ^*$,
defined recursively as follows. For $q$ $\in $ $Q$, $v$ $\in $
$\Gamma ^+$, $w$ $\in $ $\Sigma^*$, and $b$ $\in $ $\Sigma$
\begin{center}
    $\delta ^{*}(q, \lambda , v)=$$\left\{
                                      \begin{array}{ll}
                                        \delta ^{*}(\delta _Q(q, \lambda , v),
\lambda , \delta _{\Gamma ^*}(q, \lambda , v)), & \hbox{if $\delta (q, \lambda , v)\neq \perp $;} \\
                                        (q, v), & \hbox{otherwise.}
                                      \end{array}
                                    \right.$
\end{center}
\begin{center}
     $\delta ^{*}(q, wb, v)=$$\left\{
                               \begin{array}{ll}
                                 \delta ^*(\delta (\delta ^*(q, w, v), b), \lambda ), & \hbox{if
$\delta ^*(q, w, v)\neq \perp $ and $\delta (\delta ^*(q, w, v), b)\neq \perp $;} \\
                                 \perp , & \hbox{otherwise.}
                               \end{array}
                             \right.$
\end{center}

That is, $\lambda $-rules are inside the definition of $\delta
^{*}$. We abbreviate $\delta ^{*}$ to $\delta $, and $\delta
(q_0,w,z_0)$ to $\delta (w)$. We define the {\itshape output} from
state $q$ on input $w \in \Sigma ^*$ with $z \in \Gamma ^*$ on the
top of the stack by the recursion $\nu (q,\lambda ,z)= \lambda ,$
\begin{center}
    $\nu (q,wb,z)=\nu (q,w,z)\nu (\delta _Q(q,w,z), b, \delta _{\Gamma ^*}(q,w,z)).$
\end{center}
The {\itshape output} of the compressor $C$ on input $w$ $\in $
$\Sigma ^*$ is the string $C(w)=\nu (q_0,w,z_0)$.

The input of an information-lossless compressor can be reconstructed
from the output and the final state reached on that input.

\begin{definition} A PDC $C=(Q, \Sigma , \Gamma , \delta , \nu
, q_0, z_0)$ is {\itshape information-lossless} ($IL$) if the
function
\begin{center}
    $\Sigma ^*\rightarrow \Sigma ^*\times Q$
\end{center}
\begin{center}
    $w\rightarrow (C(w),\delta _Q(w))$
\end{center}
is one-to-one. An {\itshape information-lossless pushdown
compressor} ($ILPDC$) is a PDC that is IL.
\end{definition}
Intuitively, a PDC {\itshape compresses} a string $w$ if $|C(w)|$ is
significantly less than $|w|$. Of course, if $C$ is $IL$, then not
all strings can be compressed. Our interest here is in the degree
(if any) to which the prefixes of a given sequence $S$ $\in $
$\Sigma ^\infty $ can be compressed by an ILPDC.

\begin{definition} \label{def_ratio}If $C$ is a PDC and $S$
$\in $ $\Sigma ^\infty $, then the {\itshape compression ratio} of
$C$ on $S$ is
\begin{center}
    $\roC(S)=\liminf\limits_{n\rightarrow \infty }\frac{|C(S[0..n-1])|}{n\log_2(|\Sigma|)}$
\end{center}
\end{definition}
\begin{definition} The {\itshape pushdown compression ratio}
of a sequence $S$ $\in $ $\Sigma ^\infty $ is
\begin{center}
    $\rho _{PD}(S)=$ $\inf \{\roC(S)\mid \mbox{ C is an ILPDC}\}$
\end{center}
\end{definition}
We can consider dual concepts $R_C$ and $R_{PD}$ by replacing
$\liminf$ with $\limsup$ in the previous definition.

\subsection{Endmarkers and pushdown compression}\label{subsec_endmarkers}

Two possibilities occur when dealing with transducers on finite
words: should the end of the input be marked with a particular
symbol \# or not? As we will see, this is a rather subtle question.
First remark that both approaches are natural: on the one hand,
usual finite state or pushdown \emph{acceptors} do not know (and do
not need to know) when they reach the end of the word; on the other
hand, everyday compression algorithms usually know (or at least are
able to know) where the end of the input file takes place. For a
word $w$, we will denote by $C(w)$ the output of a transducer $C$
without endmarker, and $C(w\#)$ the output with an endmarker.

Unlike acceptors, transducers can take advantage of an endmarker:
they can indeed output more symbols when they reach the end of the
input word if it is marked with a particular symbol. This is
therefore a more general model of transducers which, in particular,
does not have the strong restriction of prefix extension: if there
is no endmarker and $C$ is a transducer, then for all words
$w_1,w_2$, $w_1\sqsubseteq w_2\Rightarrow C(w_1)\sqsubseteq C(w_2)$.
Let us see how this restriction limits the compression ratio.
\begin{lemma}\label{lem_no_endmarkers}
  Let $C$ be an IL pushdown compressor with $k$ states and working with no
  endmarker. Then on every word $w$ of size $|w|\geq k$, the compression ratio of $C$
  is $$\frac{|C(w)|}{|w|}\geq \frac{1}{2k}.$$
\end{lemma}
{\bfseries Proof.}
  Due to the injectivity condition, we can show that $C$ has to output at least
  one symbol every $k$ input symbols. Suppose on the contrary that
  there are words $t,u$, with $|u|=k$, such that $C$ does not output any
  symbol when reading $u$ on input $w=tu$. Then all the $k+1$ words $t$ and $tu[0..i]$
  for $0\leq i\leq k-1$ have the same output by $C$, and furthermore two of them
  have the same final state because there are only $k$ states. This contradicts injectivity. Thus $C$ must output at least
  one symbol every $k$ symbols, which proves the lemma.
\begin{flushright}
    $\Box $
\end{flushright}
This limitation does not occur with endmarkers, as the following
lemma shows.
\begin{lemma}\label{lem_yes_endmarkers}
  For every $k$, there exists an IL pushdown compressor $C$ with $k$ states, working with
  endmarkers, such that the compression ratio of $C$ on $0^n$ tends to $1/k^2$
  when $n$ tends to infinity, that is,
$$\lim\limits_{n\rightarrow \infty}\frac{|C(0^n)|}{n}= \frac{1}{k^2}.$$
\end{lemma}
{\bfseries Proof. [sketch]}
  On input $0^n$, our compressor outputs (roughly) $0^{n/k^2}$ as
  follows: by selecting one symbol out of each $k$ of the input word (counting
  modulo $k$ thanks to $k$ states), it pushes $0^{n/k}$ on the stack. Then at
  the end of the word, it pops the stack and outputs one symbol every
  $k$. Thus the output is $0^{n/k^2}$.

  To ensure injectivity, if the input word $w$ is not of the form $0^n$ (that
  is, if it contains a 1), then $C$ outputs $w$.
\begin{flushright}
    $\Box $
\end{flushright}
It is worth noticing that it is the injectivity condition that makes
this computation impossible without endmarkers, because one cannot
decide \textit{a priori} whether the input word contains a 1. Thus
pushdown compressors with endmarkers do not have the limitation of
Lemma~\ref{lem_no_endmarkers}. Still, as
Corollary~\ref{cor_universal} will show, pushdown compressors with
endmarkers are not universal for finite state compressors, in the
sense that a single pushdown compressor cannot be as good as any
finite state compressor.

It is open whether pushdown compressors with endmarkers are strictly
better than without, in the sense of the following question.

\noindent\textit{Open question.} Do there exist an infinite sequence
$S$, a constant $0<\alpha\leq 1$ and an IL pushdown compressor $C$
working with endmarkers, such that $\rho _C(S)<\alpha ,$ but $\rho
_{C'}(S)\geq \alpha ,$ for every $C'$ IL pushdown compressor working
without endmarkers?

In the rest of the paper we consider both variants of compression,
with and without endmarkers. We use the weakest variant for positive
results and the strongest for negative ones, therefore showing
stronger separations.


\section{Lempel-Ziv outperforms Pushdown transducers}\label{sec_LZoutperformsPD}
In this section we show the existence of an infinite sequence $S \in
\{0,1\}^{\infty}$ whose Lempel-Ziv almost everywhere compression
ratio is $0$ but for any IL pushdown compressor (even working with
endmarkers) the infinitely often compression ratio is $1$. The rough
idea is that Lempel-Ziv compresses repetitions very well, whereas,
if the repeated word is well chosen, pushdown compressors perform
very poorly. We first show the claim on Lempel-Ziv and then prove a
pumping-lemma for pushdown transducers in order to deal with the
case of pushdown compressors.

\subsection{Lempel-Ziv on periodic inputs} \label{subsec_LZ}  
The sequence we will build consists of regions where the same
pattern is repeated several times. This ensures that Lempel-Ziv
algorithm compresses the sequence, as shown by the following lemmas.

We begin with finite words: Lempel-Ziv compresses well words of the
form $tu^n$. The idea is that the dictionary remains small during
the execution of the algorithm because there are few different
subwords of same length in $tu^n$ due to the period of size $|u|$.
The statement is slightly more elaborated because we want to use it
in the proof of Theorem~\ref{LZ is OK_infinite} where we will need
to consider the execution of Lempel-Ziv on a nonempty dictionary.
\begin{lemma} \label{LZ is OK_finite}
Let $n \in \mathbb{N}$ and let $t, u, \in \Sigma ^*$, where $u \neq
\lambda$. Define $l=1+|t|+|u|$ and $w_n=tu^n$. Suppose we want to
run Lempel-Ziv on $w_n$, but possibly, another word has already been
parsed so that the dictionary of phrases is possibly not empty and
already contains $d>0$ phrases. Then we have that
$$\frac{|LZ(w_n)|}{|w_n|}\leq \frac{\sqrt{2l|w_n|}\log (d+\sqrt{2l|w_n|})}{|w_n|}.$$
\end{lemma}
This leads us to the following lemma on a particular infinite
sequence.
\begin{theorem}[LZ compressibility of repetitive sequences]\label{LZ is OK_infinite}
Let $(t_i)_{i\geq 1}$ and $(u_i)_{i\geq 1}$ be sequences of words,
where $u_i\neq \lambda, \forall i\geq 1$. Let $(n_i)_{i\geq 1}$ be a
sequence of integers. Let $S$ be the sequence defined by
$$S=t_1u_1^{n_1}t_2u_2^{n_2}t_3u_3^{n_3}\dots$$
If the sequence $(n_i)_{i\geq 1}$ grows sufficiently fast, then
$$R_{LZ}(S)=0.$$
\end{theorem}

\subsection{Pumping-lemma for injective pushdown transducers}\label{subsec_pumping-lemma} 
This section is devoted to the statement and proof of a
pumping-lemma for pushdown transducers. In the usual setting of
recognition of formal languages by pushdown automata, the
pumping-lemma comes from the equivalence between context-free
grammars and pushdown automata, see for instance
\cite{PapadimitriouLewis}. However, the proof is much less
straightforward without grammars, as is our case since we deal with
transducers and not acceptors. Moreover, there are three further
difficulties: first, we have to consider what happens at the end of
the word, after the endmarker (where the transducer can still output
symbols when emptying the stack); second, we need a lowerbound on
the size of the pumping part, that is, we need to pump on a
sufficiently large part of the word; third, we need the lemma for an
arbitrary finite family of automata, and not only one automaton. All
this makes the statement and the proof much more involved than in
the usual language-recognition framework (see the appendix for all
details).
\begin{lemma}[Pumping-lemma]\label{pumping-lemma}
Let $\mathcal{F}$ be a finite family of ILPDC. There exist two
constants $\alpha, \beta >0$ such that $\forall w$, there exist $t,
u, v \in \Sigma ^*$ such that  $w=tuv$ satisfying:\begin{itemize}
                                 \item $|u|\geq \lfloor\alpha
                                 |w|^{\beta}\rfloor$;
                                 \item $\forall C \in \mathcal{F}$,
                                 if $C(tuv)=xyz$, then $C(tu^n)=xy^n$, $\forall n \in \mathbb{N}$.
                               \end{itemize}
\end{lemma}
Taking into account endmarkers, we obtain the following corollary:
\begin{corollary}[Pumping-lemma with endmarkers]\label{cor_pumping}
  Let $\mathcal F$ be a finite family of ILPDC. There exist two constants $\alpha ,\beta >0$ such that every word
  $w$ can be cut in three pieces $w=tuv$ satisfying:
  \begin{enumerate}
  \item $|u|\geq \lfloor\alpha |w|^\beta\rfloor$;
  \item there is an integer $c\geq 0$ such that for all $C\in\mathcal F$,
    there exist five words $x,x',y,y',z$ such that for all $n\geq c$,
    $C(tu^nv\#)=xy^nzy'^{n-c}x'$.
  \end{enumerate}
\end{corollary}

  Let us state an immediate corollary concerning universality:
pushdown compressors, even with endmarkers, cannot be universal for
finite state compressors, in the sense that the compression ratio of
a particular pushdown compressor cannot be always better than the
compression ratio of every finite state compressor.
\begin{corollary}\label{cor_universal}
  Let $C$ be an IL pushdown compressor (with endmarkers). Then
  $\rho_C(0^{\infty})>0$. In particular, no pushdown compressor is universal
  for finite state compressors.
\end{corollary}
{\bfseries Proof.}
  By Corollary~\ref{cor_pumping}, there exist two integers $k,k'$, ($k'\geq 1$), a constant
  $c\geq 0$ and five words $x,x',y,y',z$ such that for all $n\geq c$,
  $C(0^k0^{k'n}\#)=xy^nzy'^{n-c}x'$. By injectivity of $C$, $y$ and $y'$
  cannot be both empty. Hence the size of the compression of $0^k0^{k'n}$ is
  linear in $n$. This proves the first assertion.

  Since for every $\epsilon>0$ there exists an IL finite state compressor $C'$ such
  that $R_{C'}(0^\infty)<\epsilon$, the pushdown compressor $C$ cannot be
  universal for finite state compressors.
\begin{flushright}
    $\Box $
\end{flushright}

\subsection{A pushdown incompressible sequence}\label{subsec_PDC}  
We now show that some sequences with repetitions cannot be
compressed by pushdown compressors. We start by analyzing the
performance of PDC on the factors of a Kolmogorov-random word. This
result is valid even with endmarkers.

\begin{lemma}\label{PD_bad}
For every $\mathcal{F}$ finite family of $ILPDC$ with $k$ states and
for every constant $\epsilon>0$, there exists $M_{\mathcal{F},
\epsilon} \in \mathbb{N}$ such that, for any Kolmogorov random word
$w=tu$, if $|u|\geq M_{\mathcal{F}, \epsilon}\log|w|$ then the
compression ratio for $C \in \mathcal{F}$ of $u$ on input $w$ is
$$\frac{|C(tu)|-|C(t)|}{|u|} \geq 1-\epsilon.$$
\end{lemma}
We can now build an infinite sequence of the form required in
Theorem~\ref{LZ is OK_infinite} that cannot be compressed by bounded
pushdown automata. The idea of the proof is as follows: by
Corollary~\ref{cor_pumping}, in any word $w$ we can repeat a big
part $u$ of $w$ while ensuring that the behaviour of the transducer
on every copy of $u$ is the same. If $u$ is not compressible, the
output will be of size almost $|u|$, therefore with a large number
of repetitions the compression ratio is almost $1$.
\begin{theorem}[A pushdown incompressible repetitive
sequence]\label{th_PDuncompres} Let $\Sigma$ be a finite alphabet.
There exist sequences of words $(t_k)_{k\geq 1}$ and $(u_k)_{k\geq
1}$, where $u_k\neq \lambda, \forall k\geq 1$, such that for every
sequence of integers $(n_k)_{k\geq 1}$ growing sufficiently fast,
the infinite string $S$ defined by
$$S=t_1u_1^{n_1}t_2u_2^{n_2}t_3u_3^{n_3}\dots$$ verifies that
$$\rho _{C}(S)=1,$$
$\forall C \in ILPDC$ (without endmarkers).
\end{theorem}
Combining it with Theorem~\ref{LZ is OK_infinite} we obtain the main
result of this section, there are sequences that Lempel-Ziv
compresses optimally on almost every prefix, whereas no pushdown
compresses them at all, even on infinitely many prefixes and using
endmarkers.
\begin{theorem}
There exists a sequence $S$ such that $$R_{LZ}(S)=0$$ and
$$\rho _{C}(S)=1$$ for any $C \in ILPDC$ (without endmarkers).
\end{theorem}

The situation with endmarkers is slightly more complicated, but
using Corollary~\ref{cor_pumping} (the pumping lemma with
endmarkers) and a similar construction as
Theorem~\ref{th_PDuncompres} we obtain the following result. Note
that we now use the limsup of the compression ratio for ILPDC with
endmarkers.
\begin{theorem}
There exists a sequence S such that
$$R_{LZ}(S)=0$$
and
$$R_C(S)=1$$
for any $C\in ILPDC$ (using endmarkers).
\end{theorem}


\section{Lempel-Ziv is not universal for
Pushdown compressors}\label{sec_LZnouniversalPD} It is well known
that LZ \cite{LemZiv78} yields a lower bound on the finite-state
compression of a sequence \cite{LemZiv78}, ie, LZ is universal for
finite-state compressors.

The following result shows that this is not true for pushdown
compression, in a strong sense: we construct a sequence $S$ that is
infinitely often incompressible by LZ, but that has almost
everywhere pushdown compression ratio less than $\frac{1}{2}$.

\begin{theorem} \label{th LZ_not_univ_PD}
For every $m\in\N$, there is a sequence $S \in \{0,1\}^\infty$ such
that $$\rho_{LZ}(S)
> 1 - \frac{1}{m}$$ and $$R_{PD}(S)\leq \frac{1}{2}.$$
\end{theorem}
The proof of this result is included in the appendix.


\section{Conclusion}\label{sec_conclusion}
The equivalence of compression ratio, effective dimension, and
log-loss unpredictability has been explored in different settings
\cite{FSD,Hitchcock:PHD, EFDAIT}. It is known that for the cases of
finite-state, polynomial-space, recursive, and constructive
resource-bounds, natural definitions of compression and dimension
coincide, both in the case of infinitely often compression, related
to effective versions of Hausdorff dimension, and that of almost
everywhere compression, matched with packing dimension. The general
matter of transformation of compressors in predictors and vice versa
is  widely studied \cite{ScuBro06}.

In this paper we have done a complete comparison of pushdown
compression and LZ-compression. It is  straightforward to construct
a prediction algorithm based on Lempel-Ziv compressor that uses
similar computing resources, and it is clear that finite-state
compression is always at least pushdown compression. This leaves us
with the natural open question of whether each pushdown compressor
can be transformed into a pushdown prediction algorithm, for which
the log-loss unpredictability coincides with the compression ratio
of the initial compressor, that is, whether the natural concept of
pushdown dimension defined in \cite{DotNic07} coincides with
pushdown compressibility. A positive answer would get pushdown
computation closer to finite-state devices, and a negative one would
make it closer to polynomial-time algorithms, for which the answer
is likely to be negative \cite{DIC}.

{\bfseries Acknowledgments.} The authors thank Victor Poupet for his
help on the proof of Lemma~\ref{pumping-lemma}.


\bibliographystyle{plain}




\newpage

\renewcommand{\thesection}{\Alph{section}}
\setcounter{section}{0} \setcounter{page}{1}


\pagestyle{plain}

 \vspace{8in}
\begin{center}
{\bf{\huge Technical Appendix}}
\end{center}

\section{Proofs of Section~\ref{subsec_endmarkers}}
{\bfseries Proof of Lemma~\ref{lem_yes_endmarkers}.} On input $0^n$,
the output of our pushdown compressor $C$ will be
$C(0^n\#)=01^i0^{\lfloor n/k^2\rfloor}1^j$, where $i+kj=n-k^2\lfloor
n/k^2\rfloor$, whereas on any other word $w$ (that is, any word that
contains a 1), the output will be $1w$. This ensures the injectivity
condition.

The compressor $C$ works as follows: while it reads zeroes, it
pushes on the stack one every $k$ zeroes. This is done with $k$
states by counting modulo $k$, the state $i$ meaning that the number
of zeroes up to now is $i$ modulo $k$.

  If the endmarker $\#$ is reached in state $i$ and only zeroes have been
  read, then $C$ first outputs $01^i$. Then it begins popping the stack symbol
  after symbol: it outputs one every $k$ zeroes by counting again modulo
  $k$. At the end of the stack, if it is in state $j$, it finally outputs
  $1^j$. Thus the output is $C(0^n\#)=01^i0^{\lfloor n/k^2\rfloor}1^j$ where
  $i+kj=n-k^2\lfloor n/k^2\rfloor$.

  Otherwise, when $C$ reads a $1$ in state $i$, it outputs one $1$ and completely pops
the stack while outputting $k$ zeroes at each pop, then outputs
$0^{i-1}1$, and then outputs the remaining part of the input word.
Thus the output is $C(w\#)=1w$ as claimed at the beginning of this
proof.

Thus $C$ is injective, has $k$ states, and verifies that
$\lim\limits_{n\rightarrow\infty}
  \frac{|C(0^n)|}{n}=\frac{1}{k^2}$.
\begin{flushright}
    $\Box $
\end{flushright}

\section{Proofs of Section~\ref{subsec_LZ}}
{\bfseries Proof of Lemma~\ref{LZ is OK_finite}.} Let us fix $n$ and
consider the execution of Lempel-Ziv algorithm on $w_n$: as it
parses the word, it enlarges its dictionary of phrases. Fix an
integer $k$ and let us bound the number of new words of size $k$ in
the dictionary. As the algorithm parses $t$ only, it can find at
most $|t|/k$ different words of size $k$. Afterwards, there is at
most one more word lying both on $t$ and the rest of $w_n$.

Then, due to its period of size $|u|$, the number of different words
of size $k$ in $u^n$ is at most $|u|$ (at most one beginning at each
symbol of $u$). Therefore we obtain a total of at most
$\frac{|t|}{k}+1+|u|$ different new words of size $k$ in $w_n$. This
total is upper bounded by $l=1+|t|+|u|$.

Therefore at the end of the algorithm and for all $k$, the
dictionary
  contains at most $l$ new words of size $k$. We can now upper bound the size of
  the compressed image of $w_n$. Let $p$ be the number of new phrases in the
  parsing made by Lempel-Ziv algorithm. The size of the compression is
  then $p\log(p+d)$: indeed, the encoding of each phrase consists in a new bit
  and a pointer towards one of the $p+d$ words of the dictionary. The only
  remaining step is thus to evaluate the number $p$ of new words in the
  dictionary.

 Let us order the words of the dictionary by increasing length and call $t_1$
  the total length of the first $l$ words (that is, the $l$ smallest words),
  $t_2$ the total length of the $l$ following words (that is, words of index
  between $l+1$ and $2l$ in the order), and so on: $t_k$ is the cumulated size
  of the words of index between $(k-1)l+1$ and $kl$. Since the sum of the size
  of all these words is equal to $|w_n|$, we have $$|w_n|=\sum_{k\geq 1} t_k,$$
  and furthermore, since for each $k$ there are at most $l=1+|t|+|u|$ new
  words of size $k$, we have $t_k\geq kl$. Thus we obtain
  $$|w_n|=\sum_{k\geq 1} t_k\geq \sum_{k=1}^{p/l}kl\geq \frac{p^2}{2l}.$$
  Hence $p$ satisfies
  $$\frac{p^2}{2l}\leq |w_n|,\mbox{ that is, }p\leq\sqrt{2l|w_n|}.$$
  The size of the compression of $w_n$ is $p\log
  (p+d)\leq \sqrt{2l|w_n|}\log (d+\sqrt{2l|w_n|})$, so
  $$\frac{|LZ(w_n)|}{|w_n|}\leq \frac{\sqrt{2l|w_n|}\log (d+\sqrt{2l|w_n|})}{|w_n|}.$$
\begin{flushright}
    $\Box $
\end{flushright}
\begin{remark} We have that
$$\frac{|LZ(w_n)|}{|w_n|}=O\Big(\frac {\log n}{\sqrt{n}}\Big).$$
\end{remark}
{\bfseries Proof of Theorem~\ref{LZ is OK_infinite}.} Let us call
$w_i=t_iu_i^{n_i}$ and $z_i=w_1w_2\dots w_i$. Without loss of
  generality, one can assume (for a technical reason that will become clear
  later) that for all $i$, $|z_{i-1}|$ is big enough so that for all $j\geq 0$,
  $$\sqrt{2l_i|t_iu_i^j|}\frac{\log(|z_{i-1}|+\sqrt{2l_i|t_iu_i^j|})}
  {|z_{i-1}t_iu_i^j|}\leq \frac{2i-2}{i(i+1)},$$
where $l_i=1+|t_i|+|u_i|$. By induction on $i$, let us show that we
can furthermore
  choose the integers $n_i$ so that for all $i$:
\begin{itemize}
  \item $\frac{|LZ(z_i)|}{|z_i|}\leq \frac {2}{i+1}$;
  \item for all $j< n_i$, $\frac{|LZ(z_{i-1}t_iu_i^j)|}{|z_{i-1}t_iu_i^j|}\leq \frac {4}{i+1}$.
  \end{itemize}
  This is clear for $i=1$. For $i>1$, the dictionary after $z_{i-1}$ has at
  most $|z_{i-1}|$ words. By Lemma~\ref{LZ is OK_finite} and by induction, the
  compression ratio of $z_i$ is
$$\frac{|LZ(z_i)|}{|z_i|}\leq \frac{2|z_{i-1}|/i+\sqrt{2l_i|w_i|}\log(|z_{i-1}|+\sqrt{2l_i|w_i|})}
  {|z_i|},$$
  where $l_i=1+|t_i|+|u_i|$. By taking $n_i$ sufficiently large, this can be
  made less or equal than $\frac{2}{i+1}$ and the first point of the induction follows.

For the second point, for the same reasons the compression ratio of
  $z_{i-1}t_iu_i^j$ is
  \begin{equation*}
  \begin{aligned}
  \frac{|LZ(z_{i-1}t_iu_i^j)|}{|z_{i-1}t_iu_i^j|}&\leq \frac{2|z_{i-1}|/i+\sqrt{2l_i|t_iu_i^j|}
    \log(|z_{i-1}|+\sqrt{2l_i|t_iu_i^j|})}{|z_{i-1}t_iu_i^j|}\\&=
    \frac{2}{i}+\sqrt{2l_i|t_iu_i^j|}\frac{\log(|z_{i-1}|+\sqrt{2l_i|t_iu_i^j|})}
  {\sqrt{|z_{i-1}t_iu_i^j|}}.
  \end{aligned}
  \end{equation*}
Since we have supposed without loss of generality that
  $$\sqrt{2l_i|t_iu_i^j|}\frac{\log(|z_{i-1}|+\sqrt{2l_i|t_iu_i^j|})}
  {|z_{i-1}t_iu_i^j|}\leq \frac{2i-2}{i(i+1)},$$
  we conclude that the compression ratio is less or equal than $\frac{4}{i+1}$ as required.
  Therefore the two points of the induction are proved.

  This enables us to conclude that $$R_{LZ}(S)=0.$$
\begin{flushright}
    $\Box $
\end{flushright}

\section{Proofs of Lemma~\ref{pumping-lemma}}
This section is devoted to the proof of Lemma~\ref{pumping-lemma}.
We first need several natural definitions.
\begin{definition}
Let $C$ be a $PDC$ and $w \in \Sigma^*$. \begin{itemize}
                                           \item The configuration of
                                           $C$ on $w$ in the $i$-th symbol is the pair $(q,
                                           Z_1\cdots Z_k)$ where $q \in Q$
                                           and $Z_1, \dots , Z_k \in
                                           \Gamma$ are the state and
                                           the stack content when
                                           $C$ has read $i$
                                           input symbols of $w$ (hence the last
                                           transition is not a $\lambda$-rule).
                                           Each configuration will
                                           be called column.
                                           \item The partial
                                           configuration of a column
                                           $(q, Z_1\cdots Z_k)$
                                           is $(q, Z_1)$, that is,
                                           the state and the top
                                           stack symbol of the
                                           configuration.
                                           \item The diagram of the
                                           evolution of $C$ on $w$
                                           is the sequence of all
                                           the successive columns of $C$ on the input $w$.
                                         \end{itemize}
\end{definition}
\begin{definition}
Let $C$ be a $PDC$, $w \in \Sigma ^*$ and $c$ a column in the
diagram of $C$ on input $w$.\begin{itemize}
                              \item The birth of $c$ is the index of
                              the column in the diagram.
                              \item The death of $c$ is the index of
                              the first column $c'$ appeared after $c$ in the
                              diagram such that its stack size is strictly smaller than the stack size of $c$.
                              \item The lifetime of $c$ $(life(c))$ is the
                              difference between its death and its
                              birth.
                            \end{itemize}
\end{definition}
\begin{definition}
Let $C$ be a $PDC$, $w \in \Sigma ^*$ and $c$ a column in the
diagram of $C$ on input $w$.\begin{itemize}
                              \item The children of $c$ are defined
                              as follows: let $Y_1, \dots , Y_j$ the
                              symbols pushed on the stack when a symbol
                              (having $c$ as configuration) is read . If $j=0$, then $c$ has no
                              child. Otherwise, if it exists, the first child $c_1$ is the first column
    in the lifetime of $c$; the second child $c_2$ is the column where $c_1$
    dies (that is, the first column where the stack is strictly smaller than
    in $c_1$) if it happens during the lifetime of $c$; the third child $c_3$
    is the column where $c_2$ dies if it happens during the lifetime of $c$,
    and so on. In particular, $c$ has at most $j$ children because the size of
    the stack in $c_{i+1}$ is strictly smaller than in $c_i$, and the death of
    a child always happens before the death of $c$.
                              \item A descendant of a column is
                              either a child or a child of a
                              descendant.
                            \end{itemize}
\end{definition}
The following definition will be useful in order to repeat a part of
the diagram (that is, to `pump').

\begin{definition}
Let $C$ be a $PDC$ and $c,d$ two columns in the diagram of $C$ on
input $w \in \Sigma^*$. Then $c$ and $d$ are said to be equivalent
if one is a descendant of the other and they have the same partial
configuration. In
  particular, if we denote by $u$ the input word between $c$ and $d$, then for
  all $i\geq 1$, after reading $u^i$ we will obtain another column $e$
  equivalent to $c$ and $d$.
\end{definition}
Moreover, we need some definitions regarding a finite family of
compressors.

\begin{definition}
Let $\mathcal{F}$ be a finite family of $PDC$.
\begin{itemize}
                 \item Let $c_1, \dots , c_{|\mathcal{F}|}$ be the
                 columns of all $C \in \mathcal{F}$ at a given
                 input symbol (that is, with the same index). Then the tuple $(c_1, \dots ,
                 c_{|\mathcal{F}|})$ is called a generalized column
                 of $\mathcal{F}$.
                 \item Let $q_1, \dots , q_{|\mathcal{F}|}$ be the
                 partial configurations of all $C \in \mathcal{F}$
                 at a given input symbol. Then the tuple $(q_1, \dots ,
                 q_{|\mathcal{F}|})$ is called a generalized
                 partial configuration (gpc for short) of $\mathcal{F}$.
                 \item Two generalized columns are equivalent if
                 for all $C \in \mathcal{F}$, the corresponding
                 columns are equivalent. In particular, the gpc are
                 the same.
                 \item The lifetime of a generalized column $(life((c_1, \dots ,
                 c_{|\mathcal{F}|})))$ is the
                 minimum of the lifetimes of the
                 corresponding columns of all $C \in \mathcal{F}$.
               \end{itemize}
\end{definition}
We can now begin the proof of Lemma~\ref{pumping-lemma}. In order to
be able to pump, we need to find two equivalent gpc. Since the
family $\mathcal F$ of transducers is finite, there is a finite
number of transition rules, therefore let $k$ be the maximum size of
a word pushed on the stack by one rule. The integer $p$ will denote
the total number of gpc, that is, the product over all $T\in\mathcal
F$ of the number of states of $T$ multiplied by the size of the
stack alphabet of $T$. Finally, $d$ will be an integer representing
a distance.

We will upperbound the size of a counterexample, that is, the size
of a word in which no pair of generalized columns at distance more
than or equal to $d$ are equivalent. More precisely, let $L(p,k,d)$
be the maximum lifetime of a generalized column during which:
\begin{itemize}
\item only $p$ distinct gpc appear;
\item the size of a word pushed on the stack by one rule is less than or equal to $k$;
\item no pair of equivalent generalized columns are at distance more than or equal to $d$.
\end{itemize}
Let us upperbound $L(p,k,d)$ using an induction.

\begin{claim}\label{claim_case1}
For all $k, d \in \mathbb{N}$, $$L(1, k, d)<d.$$
\end{claim}
{\bfseries Proof.} There is only one possible gpc. If the lifetime
of a generalized columns was greater than or equal to $d$, then
there would be two equivalent generalized columns at distance at
least $d$.
\begin{flushright}
    $\Box$
\end{flushright}
\begin{claim}\label{claim_casegeneral}
$\forall p, k, d \in \mathbb{N}$, $$L(p+1, k, d)\leq d +
|\mathcal{F}|kdL(p, k, d).$$
\end{claim}
{\bfseries Proof.} Let $(c_1, \dots , c_{|\mathcal{F}|})$ be a
generalized column of the diagram of $\mathcal{F}$ on $w$ whose gpc
is $(q_1, \dots , q_{|\mathcal{F}|})$, and let us upperbound its
lifetime. During this lifetime, the first $d$ generalized columns
can be arbitrary but the remaining ones cannot contain the gpc
$(q_1, \dots , q_{|\mathcal{F}|})$ since they are descendant of
$(c_1, \dots , c_{|\mathcal{F}|})$ (otherwise, we would obtain two
equivalent generalized columns at distance more than or equal to
$d$, which is not possible). Hence every generalized column at
distance at least $d$ from $(c_1, \dots , c_{|\mathcal{F}|})$ has
its lifetime bounded by $L(p, k, d)$.

Let $C \in \mathcal{F}$. For convenience, we count the size of the
stack relatively to that in $(c_1, \dots , c_{|\mathcal{F}|})$, i.e,
we will say that the size of the stack in $(c_1, \dots ,
c_{|\mathcal{F}|})$ is $0$. Then the stack after the first $d$
columns is of size at most $dk$, because every rule can push at most
$k$ symbols each time. We call $c_{C_1}$ the column of $C$ at
distance $d$ of $(c_1, \dots , c_{|\mathcal{F}|})$, $c_{C_2}$ the
column of $C$ whose birth coincides with the death of $c_{C_1}$,
$c_{C_3}$ the column whose birth coincides with the death of
$c_{C_2}$, and so on. The intervals between $c_{C_1}$ and $c_{C_2}$,
between $c_{C_2}$ and $c_{C_3}$, etc, will be referred to as
`disjoint intervals'. Remark that the size of the stack in
$c_{C_{i+1}}$ is less than the size in $c_{C_i}$ (because the top
symbol in $c_{C_i}$ has been popped). Hence the number of these
columns is at most $dk$ (because the size of the stack in $c_{C_1}$
was at most $dk$).

Back to the whole family $\mathcal{F}$, this enables us to bound the
number of disjoint generalized intervals at distance at least $d$
from $(c_1, \dots , c_{|\mathcal{F}|})$. Indeed, since every
compressor has at most $dk$ disjoint intervals, then there are at
most $|\mathcal{F}|kd$ disjoint generalized intervals at distance at
least $d$ from $(c_1, \dots , c_{|\mathcal{F}|})$ (because the
definition of the lifetime of a generalized column is the minimum of
the lifetimes of each column). Since each of them has its lifetime
bounded by $L(p, k, d)$, this proves the claim.
\begin{flushright}
    $\Box $
\end{flushright}
\begin{claim}\label{claim_end}
$\forall p, k, d \in \mathbb{N}$, $$L(p, k,
d)<(|\mathcal{F}|kd)^{p+1}.$$
\end{claim}
{\bfseries Proof.} Iterating claim \ref{claim_casegeneral}, we
obtain that
\begin{equation*}
\begin{aligned}
L(p, k, d)&<d\sum \limits _{i=0}^{p-1} (|\mathcal{F}|kd)^i
=d\frac{1-(|\mathcal{F}|kd)^p}{1-|\mathcal{F}|kd}\\
&=d\frac{(|\mathcal{F}|kd)^p-1}{|\mathcal{F}|kd-1} \leq
(|\mathcal{F}|kd)^{p+1}.\end{aligned}\end{equation*}
\begin{flushright}
    $\Box$
\end{flushright}
{\bfseries Proof of Lemma \ref{pumping-lemma}.} For the finite
family $\mathcal{F}$, we have fixed constants $p$ and $k$. Now, take
a word $w$ and let $$d=\Big \lfloor
\frac{|w|^{\frac{1}{p+1}}}{k|\mathcal{F}|}\Big \rfloor.$$ Then, by
using Claim \ref{claim_end}, we obtain that $$L(p, k, d)<|w|,$$ so
on $w$ there are two equivalent generalized columns at distance more
than or equal to $d$. We call $u$ the part of the input read between
these two generalized columns ($|u|\geq d$, that is, $|u|\geq
\lfloor \alpha |w|^{\beta}\rfloor$ for some constants
$\alpha,\beta>0$), $t$ the part of the input previous to $u$ and $v$
the part following $u$. We have that
$$C(tuv)=xyz$$ $\forall C \in
\mathcal{F}$. Then, $C (tu^n)=xy^n$ $\forall n \in \mathbb{N}$ and
$\forall C \in \mathcal{F}$. This concludes the proof of
Lemma~\ref{pumping-lemma}.

\begin{flushright}
    $\Box $
\end{flushright}

{\bfseries Proof of Corollary~\ref{cor_pumping}.}
 The proof is very similar as that of Lemma~\ref{pumping-lemma} but we have to take into account the computation
  after the endmarker \#. Remark that an ILPDC behaves like a finite state
  transducer after \#, because it can only read the stack and pop its top
  symbol. We therefore only have to take into account the state of the
  transducer in this phase. Thus, instead of merely looking for two equivalent
  generalized partial configurations as in the proof of
  Lemma~\ref{pumping-lemma}, we are looking for two equivalent generalized
  partial configurations for which the ``corresponding state'' after \#\ while
  popping the stack is the same. The argument then goes through by considering
  pairs (gpc,state), but now $v$ plays a role as being the word influencing
  the topmost symbols of the stack.
\begin{flushright}
    $\Box $
\end{flushright}

\section{Proofs of Section~\ref{subsec_PDC}}
Lemma~\ref{PD_bad} follows directly from this lemma.
\begin{lemma}\label{size_u}
Let $w \in \Sigma ^*$ be a Kolmogorov random word (that is, $w$ is
such that $K(w)\geq |w|$) and $C$ be an $ILPDC$ with $k$ states. Let
$|C|$ be the size of its encoding (i.e. the complete description of
$C$). For every $t, u \in \Sigma ^*$ such that $w=tu$
$$|C(tu)|-|C(t)|\geq |u|-|C|-\log k -2\log |w|-2\log |C|-2\log \log k-3.$$
\end{lemma}
{\bfseries Proof of Lemma~\ref{size_u}.} Given $w$, since $C$ is
$IL$, $w$ can be recovered from $t$, $C$, $\nu (u)$ and the final
state $q$ of $C$ on input $w$. If we encode a tuple $(x_1, \dots ,
x_m)$ as $$1^{\lceil \log n_1\rceil }0n_1x_11^{\lceil \log n_2\rceil
}0n_2x_2\dots 1^{\lceil \log n_{m-1}\rceil }0n_{m-1}x_{m-1}x_m,$$
where $n_i=|x_i|$ is in binary, then encoding the $4$-tuple $(t, C,
\nu (u), q)$ takes size
$$2\lceil \log |t|\rceil +1+|t|+2\lceil \log |C|\rceil +1+|C|+2\lceil \log \lceil
\log k\rceil \rceil +1+\lceil \log k\rceil +|\nu (u)|,$$ where $k$
is the number of states of $C$. We therefore obtain
$$|w|\leq K(w)\leq 2\log |t|+|t|+2\log |C|+|C|+2\log \log k+\log k+|\nu (u)|+O(1)$$ and the lemma follows.
\begin{flushright}
    $\Box $
\end{flushright}

{\bfseries Proof of Theorem~\ref{th_PDuncompres}.}
  For all $k$, let $\mathcal{F}_k$ be the (finite) family of ILPDC with at
  most $k$ states and with a stack alphabet of size $k$. Let $\alpha_k$ and
  $\beta_k$ be the constants given by Corollary~\ref{cor_pumping} for the
  family $\mathcal{F}_k$. Let $M_k=M_{\mathcal{F}_k,1/k}$ as given by
  Lemma~\ref{PD_bad}.

  For all $k$, take a Kolmogorov-random word $w_k$ of size big enough so that
  $\lfloor\alpha_k|w_k|^{\beta_k}\rfloor>M_k\log|w_k|$. By
  Corollary~\ref{cor_pumping}, $w_k$ can be cut in three pieces
  $w_k=t_ku_kv_k$ such that $|u_k|>M_k\log|w_k|$. For all $k$, define $n'_k$
  to be the least integer so that
  $(1-1/k)|u_k^{n'_k}|>(1-2/k)|t_ku_k^{n'_k}|$. We claim that the sequence
  $$S=t_1u_1^{n_1}t_2u_2^{n_2}t_3u_3^{n_3}\dots$$
  fulfills the requirements of the lemma as soon as the sequence $(n_k)_{k\geq
    1}$ is chosen so that for all $k$, we have
  \begin{enumerate}
  \item $n_k\geq n'_k$;
  \item $|t_{k+1}u_{k+1}^{n_{k+1}}| > |t_1u_1^{n_1}\dots t_ku_k^{n_k}|$;
  \item $|t_ku_k^{n_k}|>k|t_{k+1}(u_{k+1})^{n'_{k+1}}|$.
  \end{enumerate}
  Indeed, condition 1 ensures that no automaton with $\leq k$ states and of
  stack alphabet of size $k$ can compress $t_ku_k^{n_k}$ with a ratio better
  than $1-2/k$; then condition 2 shows that asymptotically the compression
  ratio of the whole prefix $t_1u_1^{n_1}\dots t_ku_k^{n_k}$ tends to 1 as $k$
  tends to infinity. Finally, the third condition ensures that for all $i$,
  the compression ratio of $t_ku_k^{n_k}t_{k+1}n_{k+1}^i$ also tends to 1 as
  $k$ tends to infinity. As a whole, for every automaton, the liminf of the
  compression ratio on $S$ is at least 1.
\begin{flushright}
    $\Box $
\end{flushright}

\section{Proof of Theorem~\ref{th LZ_not_univ_PD}}
In this proof we work with the binary alphabet, the general case can
be proven similarly.
{\bfseries Proof.}
 Let $m\in\N$, and let
    $k=k(m), v=v(m), v'=v'(m)$ be  integers to be determined later.
        For any integer $n$, let $T_n$ denote the set of  strings $x$ of size $n$ such that
        $1^j$ does not appear in $x$, for every $j\geq k$.
        Since $T_n$ contains $\bool{k-1}\times \{ 0 \} \times \bool{k-1}\times \{ 0 \} \ldots $
        (i.e. the set of strings whose every $k$th bit is zero),
        it follows that
        $|T_n|\geq 2^{an}$, where $a=1-1/k$.

        \begin{remark} \label{r.extension}
            For every string $x\in T_n$ there is a string $y\in T_{n-1}$ and a bit $b$ such that $yb=x$.
        \end{remark}

        Let $A_n = \{a_1,\ldots a_u\}$ be the set of palindromes in $T_n$. Since fixing the $n/2$ first bits of a palindrome (wlog $n$ is even)
        completely determines it, it follows that $|A_n| \leq 2^{\frac{n}{2}}$.
        Let us separate the remaining strings in $T_n-A_n$ into $v$ pairs of sets
        $X_{n,i} = \{x_{i,1},\ldots x_{i,t}\}$ and $Y_{n,i} = \{y_{i,1},\ldots y_{i,t}\}$ with
    $t=\frac{|T_n-A_n|}{2v}$,
        $(x_{i,j})^{-1}=y_{i,j}$ for every $1\leq j \leq t$ and $1\leq i \leq v$,
    $x_{i,1}, y_{i,t}$ start with a zero. For convenience we write $X_{i}$ for $X_{n,i}$.

        We construct $S$ in stages.
    Let $f(k)=2k$ and $f(n+1)=f(n)+v+1$. Clearly $$ n^2>f(n)>n.$$ For $n \leq k-1$,
        $S_n$ is an enumeration of all strings of size $n$ in lexicographical order.
        For $n\geq k$,
        $$S_n = a_1 \ldots a_u \  1^{f(n)} \ x_{1,1} \ldots x_{1,t} \ 1^{f(n)+1} \ y_{1,t} \ldots y_{1,1}
    \ldots
    x_{v,1} \ldots x_{v,t} 1^{f(n)+v} y_{v,t} \ldots y_{v,1}$$
        i.e. a concatenation of all strings in $A_n$ (the $A$ zone of $S_n$) followed by a flag of $f(n)$ ones,
        followed by the concatenations of all strings in the  $X_i$ zones and $Y_i$ zones, separated
    by flags of increasing length. Note that the $Y_i$ zone is exactly the $X_i$ zone written in reverse
    order.
    \noindent
        Let $$S=S_1 S_2 \ldots S_{k-1} \ 1^k \ 1^{k+1} \ \ldots 1^{2k-1} \ S_{k} S_{k+1} \ldots $$
        i.e. the concatenation of the $S_j$'s with some extra flags between $S_{k-1}$ and $S_k$.
        We claim that the parsing of $S_n$ ($n\geq k$) by LZ, is as follows:
        $$a_1, \ldots , a_u, \  1^{f(n)}, \ x_{1,1}, \ldots, x_{1,t}, \ 1^{f(n)+1}, \ y_{1,t}, \ldots, y_{1,1},
    \ldots,
    x_{v,1}, \ldots, x_{v,t}, 1^{f(n)+v}, y_{v,t}, \ldots, y_{v,1} .$$
        Indeed after $S_1, \ldots S_{k-1} \ 1^k \ 1^{k+1} \ \ldots 1^{2k-1}$, LZ has parsed every
        string of size $\leq k-1$ and the flags $1^k \ 1^{k+1} \ \ldots 1^{2k-1}$. Together with Remark \ref{r.extension},
        this guarantees that LZ parses $S_n$ into phrases that are exactly all the strings in $T_n$ and
        the $v+1$ flags $1^{f(n)},\ldots,1^{f(n)+v}$.

        Let us compute the compression ratio $\rho_{LZ} (S)$.
        Let $n,i$ be integers. By construction of $S$, LZ encodes every phrase in $S_i$ (except flags),
          by a phrase in $S_{i-1}$ (plus a bit).
        Indexing a phrase in $S_{i-1}$ requires a codeword of length at least logarithmic in the number of phrase parsed
        before, i.e. $\log (P(S_1 S_2 \ldots S_{i-2}))$.
        Since $P(S_i)\geq |T_i| \geq 2^{ai}$, it follows
        $$
        P(S_1 \ldots S_{i-2}) \geq \sum^{i-2}_{j=1}2^{aj} = \frac{2^{a(i-1)} -2^a}{2^a-1} \geq b 2^{a(i-1)}
        $$
        where $b=b(a)$ is arbitrarily close to $1$.
        Letting $t_i=|T_i|$, the number of bits output by LZ on $S_i$ is at least
        \begin{align*}
        P(S_i) \log P(S_1\ldots S_{i-2})
        &\geq t_i \log b 2^{a(i-1)}\\
        &\geq c t_i(i-1)
        \end{align*}
        where $c=c(b)$ is arbitrarily close to $1$.
        Therefore
        $$
        |LZ(S_1 \ldots S_n)| \geq  \sum_{j=1}^{n} c t_j(j-1)
        $$
        Since
    $$|S_1 \ldots S_n| = |S_1 \ldots S_{k-1} 1 \ldots 1|+|S_{k}\ldots S_{n}|
    \leq 2^{3k} + \sum_{j=k}^{n} (j t_j + (v+1)(f(j)+v))  $$
    and
    $|LZ(S_1\ldots S_n)| \geq 0+\sum_{j=k}^{n}ct_j(j-1)$,
    the compression ratio is given by
        \begin{align}
        \rho_{LZ}(S_1 \ldots S_n)   &\geq
    c\frac{\sum_{j=k}^{n} t_j(j-1) }{2^{3k} + \sum_{j=k}^{n} (j t_j + (v+1)(f(j)+v))}\\
        &= c - c \frac{2^{3k} + \sum_{j=k}^{n} (j t_j + (v+1)(f(j)+v) -t_j(j-1))  }
    {2^{3k} + \sum_{j=k}^{n} (j t_j + (v+1)(f(j)+v))} \\
    &= c - c\frac{2^{3k} + \sum_{j=k}^{n} (t_j + (v+1)(f(j)+v))  }
    {2^{3k} + \sum_{j=k}^{n} (j t_j + (v+1)(f(j)+v))  } \label{e.secondterm}
        \end{align}
        The second term in Equation \ref{e.secondterm} can be made arbitrarily small for $n$ large enough:
        Let $k < M\leq n$, we have
        \begin{align*}
        \sum_{j=k}^{n} j t_j &\geq \sum_{j=k}^{M} j t_j + (M+1)\sum_{j=M+1}^{n}  t_j\\
        &= \sum_{j=k}^{M} j t_j + M\sum_{j=M+1}^{n} t_j+ \sum_{j=M+1}^{n}  t_j\\
          &\geq \sum_{j=k}^{M} j t_j + M\sum_{j=M+1}^{n} t_j+ \sum_{j=M+1}^{n}  2^{aj}\\
          &\geq  \sum_{j=k}^{M} j t_j + M\sum_{j=M+1}^{n} t_j+ 2^{an}\\
        \end{align*}
    We have
    $$
    2^{an} \geq M [ 2^{3k} +\sum_{j=k}^{M} t_j + (v+1) \sum_{j=k}^{n} (f(j)+v) ]
    $$
    for $n$ large enough, because $f(j)<j^2$.
    Hence
    $$
    c\frac{2^{3k} + \sum_{j=k}^{n} (t_j + (v+1)(f(j)+v))  }
    {2^{3k} + \sum_{j=k}^{n} (j t_j + (v+1)(f(j)+v))  }
    \geq
    c\frac{2^{3k} + \sum_{j=k}^{n} (t_j + (v+1)f(j)+v)  }
    {M[ 2^{3k} + \sum_{j=k}^{n} (t_j + (v+1)(f(j)+v))  ]} = \frac{c}{M}
    $$
    i.e.
        $$
        \rho_{LZ}(S_1 \ldots S_n)   \geq c - \frac{c}{M}
        $$
        which by definition of $c,M$ can be made arbitrarily close to $1$ by choosing $k$ accordingly, i.e
        $$
        \rho_{LZ}(S_1 \ldots S_n)   \geq 1- \frac{1}{m}.
        $$

        Let us show that $R_{PD}(S)\leq \frac{1}{2}$. Consider the following ILPD compressor $C$.
        On any of the zones $A,X_i$ and the flags, $C$ outputs them bit by bit; on $Y_i$ zones, $C$
          outputs 1 bit for every $v'$ bits of input. For the stack: $C$ on  $S_n$ cruises through the $A$ zone until the
            first flag, then starts pushing the whole $X_1$
            zone onto its stack until it hits the second flag. On $Y_1$, $C$ outputs a $0$ for every $v'$ bits of input,
            pops on symbol from the stack for every bit of input, and cruises through $v'$ counting states, until the stack is
            empty (i.e. $X_2$ starts).
            $C$  keeps doing the same for each pair $X_i,Y_i$ for every
            $2\leq i \leq v$. Therefore at any time, the number of bits of $Y_i$ read so far is equal to
            $v'$ times the number of bits output on the $Y_i$ zone plus the index of the current counting state.
            On the $Y_i$ zones, $C$ checks that every bit of $Y_i$ is equal to the bit it pops from the stack;
            if the test fails, $C$ enters an error state and outputs every bit it reads from then on (this guarantees
            IL on sequences different from $S$).
            This together with the fact that the $Y_i$ zone is exactly the $X_i$ zone written in reverse order,
            guarantees that $C$ is IL. Before giving a detailed
        construction of $C$, let us compute the upper bound it yields on $R_{PD}(S)$.

    \begin{remark}\label{r.martingale.capital}
        For any $j\in\N$, let $p_j=C(S[1\ldots j])$ be the output of $C$ after reading $j$ bits of $S$.
        Is it easy to see that the ratio $\frac{|p_j|}{|S[1\ldots j]|}$ is maximal
        at the end of a flag following an $X_i$ zone, (since the flag is followed by a $Y_i$ zone,
        on which $C$ outputs a bit for every $v'$ input bits).
    \end{remark}

    Let $1\leq t \leq v$. We compute the ratio $\frac{|p_j|}{|S[1\ldots j]|}$ inside zone $S_n$
    on the last bit of the flag following $X_{t+1}$. At this location (denoted $j_0$), $C$ has output
    \begin{align*}
    |p_{j_0}| &\leq 2^{3k} + \sum_{j=k}^{n-1}[j|A_j| + (v+1)(f(j)+v)+\frac{j}{2}|T_j-A_j|(1+\frac{1}{v'})]
                + n|A_n|+(v+1)(f(n)+v) \\&+ \frac{n}{2v}|T_n-A_n|(t+1+\frac{t}{v'})\\
                &\leq 2^{pn} + \sum_{j=k}^{n-1}[\frac{j}{2}|T_j|(1+\frac{1}{v'})]
                + \frac{n}{2v}|T_n|(t+1+\frac{t}{v'})
    \end{align*}
    where $p>\frac{1}{2}$ can be made arbitrarily close to $\frac{1}{2}$.

    The number of bits of $S$ at this point is
    \begin{align*}
        |S[1\ldots j_0]| &\geq \sum_{j=k}^{n-1}j|T_j|
                + n|A_n| + \frac{n}{v}|T_n-A_n|(t+\frac{1}{2})\\
                &\geq  \sum_{j=k}^{n-1}j|T_j|
                + \frac{n}{v}|T_n|(t+\frac{1}{4})
    \end{align*}

    Hence by Remark \ref{r.martingale.capital}
    \begin{align*}
        \liminf_{n\rightarrow\infty} \frac{|p_n|}{|S[1\ldots n]|}
      &\leq \liminf_{n\rightarrow\infty} \frac{2^{pn} + \sum_{j=k}^{n-1}[\frac{j}{2}|T_j|(1+\frac{1}{v'})]
                + \frac{n}{2v}|T_n|(t+1+\frac{t}{v'})}{\sum_{j=k}^{n-1}j|T_j|+ \frac{n}{v}|T_n|(t+\frac{1}{4})}\\
       &= \liminf_{n\rightarrow\infty}
        [
        \frac{2^{pn}}{\sum_{j=k}^{n-1}j|T_j|+ \frac{n}{v}|T_n|(t+\frac{1}{4})}
        + \frac{1}{2}\frac{\sum_{j=k}^{n-1}j|T_j| + \frac{n|T_n|}{v}(t+\frac{1}{4})}
            {\sum_{j=k}^{n-1}j|T_j|+ \frac{n}{v}|T_n|(t+\frac{1}{4})}
        \\&+ \frac{1}{2v'}\frac{\sum_{j=k}^{n-1}j|T_j|}{\sum_{j=k}^{n-1}j|T_j|+ \frac{n}{v}|T_n|(t+\frac{1}{4})}
        + \frac{n|T_n|}{2v}\frac{\frac{t}{v'}+\frac{3}{4}}{\sum_{j=k}^{n-1}j|T_j|+ \frac{n}{v}|T_n|(t+\frac{1}{4})}
        ]
   \end{align*}

    Since $\sum_{j=k}^{n-1}j|T_j| \geq (n-1)|T_{n-1}|\geq (n-1)\frac{|T_n|}{2}$,
    we have
    \begin{align*}
    \sum_{j=k}^{n-1}j|T_j| + \frac{n}{v}|T_n|(t+\frac{1}{4})
    &\geq \frac{n-1}{2}|T_n|+\frac{n}{v}|T_n|(t+\frac{1}{4})\\
    &= \frac{n|T_n|}{2v}(v-\frac{v}{n}+2t+\frac{1}{2}).
    \end{align*}
    Therefore
    \begin{align*}
    \liminf_{n\rightarrow\infty} \frac{2^{pn}}{\sum_{j=k}^{n-1}j|T_j|+ \frac{n}{v}|T_n|(t+\frac{1}{4})}
    &\leq \liminf_{n\rightarrow\infty} \frac{2^{pn}}{\frac{(n-1)}{2}|T_n|}\\
    &\leq \liminf_{n\rightarrow\infty} \frac{2^{pn}}{2^{an}} = 0
    \end{align*}
    and
    $$
        \frac{1}{2v'}\frac{\sum_{j=k}^{n-1}j|T_j|}{\sum_{j=k}^{n-1}j|T_j|+ \frac{n}{v}|T_n|(t+\frac{1}{4})}
        \leq \frac{1}{2v'}
    $$
    which is arbitrarily small by choosing $v'$ accordingly, and
    $$
    \frac{n|T_n|}{2v}\frac{\frac{t}{v'}+\frac{3}{4}}{\sum_{j=k}^{n-1}j|T_j|+ \frac{n}{v}|T_n|(t+\frac{1}{4})}
    \leq \frac{\frac{t}{v'}+\frac{3}{4}}{v-\frac{v}{n}+2t+1}
    $$
    which is arbitrarily small by choosing $v$ accordingly.
    Thus
   $$
        R_{PD}(S)
        = \liminf_{n\rightarrow\infty} \frac{|p_n|}{|S[1\ldots n]|}  \leq \frac{1}{2}.
    $$

        Let us give a detailed description of $C$. Let $Q$ be the following set of states:
        \begin{itemize}
            \item   The start state $q_0$, and $q_1,\ldots q_w \ $ the ``early'' states that will count up to
                    $$w=|S_1 S_2 \ldots S_{k-1} \ 1^k \ 1^{k+1} \ \ldots 1^{2k-1}|.$$
            \item   $q^a_0, \ldots, q^a_k \quad$ the $A$ zone states that cruise through the $A$ zone until the first flag.
            \item   $q^{f}_{j} \quad$  the $j$th flag state, ($j=1,\ldots,v+1$)
            \item   $q^{X_j}_0, \ldots, q^{X_j}_k \quad$ the $X_j$ zone states that cruise through the $X_j$
            zone, pushing every bit on the stack,
                    until the $(j+1)$-th flag is met, ($j=1,\ldots,v$).
            \item   $q^{Y_j}_1, \ldots, q^{Y_j}_{v'} \quad$ the $Y_j$ zone states that cruise through the $Y_j$
            zone, poping 1 bit  from the stack (per input bit) and comparing it to the input bit,
             until the stack is empty, ($j=1,\ldots,v$).
                \item   $q^{r,j}_0, \ldots, q^{r,j}_k \quad$ which after the $j$th flag is detected,
            pop $k$ symbols from the stack that were
                    erroneously pushed while reading the $j$th flag, ($j=2,\ldots,v+1$).
            \item $q_{e}\quad$ the error state, if one bit of $Y_i$ is not equal to the content of the stack.
        \end{itemize}
        Let us describe the transition function $\delta :Q \times \bool{} \times\bool{}\rightarrow Q\times\bool{}$.
        First $\delta$ counts until $w$ i.e. for $i=0,\ldots w-1$
        $$
            \delta(q_i,x,y) = (q_{i+1},y) \quad \text{ for any } x,y
        $$
        and after reading $w$ bits, it enters in the first $A$ zone state, i.e. for any $x,y$
        $$\delta(q_w,x,y) = (q^a_0,y).$$
        Then $\delta$ skips through $A$ until the string $1^k$ is met, i.e. for $i=0,\ldots k-1$ and any $x,y$
        $$
            \delta(q^a_i,x,y) =
        \begin{cases}
        (q^a_{i+1},y) &\text{ if } x=1\\
        (q^a_{0},y) &\text{ if } x=0\\
        \end{cases}
        $$ and
        $$
        \delta(q^a_k,x,y) = (q^{f}_1,y).
        $$
        Once $1^k$ has been seen, $\delta$ knows the first flag has started, so it skips
        through the flag until a zero is met, i.e. for every $x,y$
        $$
            \delta(q^{f}_1,x,y) =
        \begin{cases}
        (q^{f}_1,y) &\text{ if } x=1\\
        (q^{X_1}_{0},0y) &\text{ if } x=0\\
        \end{cases}
        $$
        where state $q^{X_1}_0$ means that the first bit of the $X_1$ zone (a zero bit) has been read, therefore $\delta$ pushes a zero.
        In the $X_1$ zone, delta pushes every bit it sees until it reads a sequence of $k$ ones, i.e until the start of the second flag, i.e
        for $i=0,\ldots k-1$ and any $x,y$
        $$
            \delta(q^{X_1}_i,x,y) =
        \begin{cases}
        (q^{X_1}_{i+1},xy) &\text{ if } x=1\\
        (q^{X_1}_{0},xy) &\text{ if } x=0\\
        \end{cases}
        $$
        and
        $$
        \delta(q^{X_1}_k,x,y) = (q^{r,2}_0,y).
        $$
        At this point, $\delta$ has pushed all the $X_1$ zone on the stack, followed by
        $k$ ones. The next step is to pop $k$ ones,
        i.e
        for $i=0,\ldots k-1$ and any $x,y$
        $$
            \delta(q^{r,2}_i,x,y) = (q^{r,2}_{i+1},\lambda)
        $$
        and
        $$
        \delta(q^{r,2}_k,x,y) = (q^{f}_2,y).
        $$
        At this stage, $\delta$ is still in the second flag (the second flag is always bigger than $2k$)
        therefore it keeps on reading ones until a zero (the first bit of the $Y$ zone) is met. For any $x,y$
        $$
            \delta(q^{f}_2,x,y) =
        \begin{cases}
        (q^{f}_2,y) &\text{ if } x=1\\
        (q^{Y_1}_1,\lambda) &\text{ if } x=0 .
        \end{cases}
        $$
        On the last step, $\delta$ has read the first bit of the $Y_1$ zone, therefore it pops it. At this stage,
        the stack  exactly contains the $X_1$ zone written in reverse order (except the first bit),
        $\delta$ thus uses its stack to check that what follows is really the $Y_1$ zone. If it is not the case,
            it enters $q_e$. While cruising through $Y_1$, $\delta$  counts with period $v'$. Thus
            for $i=1,\ldots v'-1$ and any $x,y$
        $$
            \delta(q^{Y_1}_i,x,y) =
        \begin{cases}
        (q^{Y_1}_{i+1},\lambda) &\text{ if } x=y\\
        (q_e,\lambda) &\text{ otherwise } \\
        \end{cases}
        $$
        and
            $$
            \delta(q^{Y_1}_{v'},x,y) =
        \begin{cases}
        (q^{Y_1}_{1},\lambda) &\text{ if } x=y\\
        (q_e,\lambda) &\text{ otherwise } \\
        \end{cases}
        $$

         Once the stack is empty,
        the $X_2$ zone begins. Thus, for any $x,y$, $1\leq i \leq v'$
        $$\delta(q^{Y_1}_i,x,z_0) =
        \begin{cases}
        (q^{X_2}_1,1z_0) &\text{ if } x=1\\
        (q^{X_2}_0,0z_0) &\text{ if } x=0 .
        \end{cases}
        $$
    Then for $2\leq j \leq v$ and $0\leq i \leq k-1$, the behaviour is the same, i.e.
    $$
            \delta(q^{X_j}_i,x,y) =
        \begin{cases}
        (q^{X_j}_{i+1},xy) &\text{ if } x=1\\
        (q^{X_j}_{0},xy) &\text{ if } x=0\\
        \end{cases}
        $$and
        $$
        \delta(q^{X_j}_k,x,y) = (q^{r,j+1}_0,y).
        $$
    At this stage we reached the end of the $(j+1)$th flag , therefore we quit $k$ bits from the stack.
    $$
            \delta(q^{r,j+1}_i,x,y) = (q^{r,j+1}_{i+1},\lambda)
        $$
        and
        $$
        \delta(q^{r,j+1}_k,x,y) = (q^{f}_{j+1},y).
        $$
    At this stage $\delta$ is in the $(j+1)$ th flag, thus:
    $$
            \delta(q^{f}_{j+1},x,y) =
        \begin{cases}
        (q^{f}_{j+1},y) &\text{ if } x=1\\
        (q^{Y_j}_1,\lambda) &\text{ if } x=0 .
        \end{cases}
        $$
    Next the $Y_j$ zone has been reached, so for $i=1,\ldots v'-1$ and any $x,y$
        $$
            \delta(q^{Y_j}_i,x,y) =
        \begin{cases}
        (q^{Y_j}_{i+1},\lambda) &\text{ if } x=y\\
        (q_e,\lambda) &\text{ otherwise } \\
        \end{cases}
        $$
        and
            $$
            \delta(q^{Y_j}_{v'},x,y) =
        \begin{cases}
        (q^{Y_j}_{1},\lambda) &\text{ if } x=y\\
        (q_e,\lambda) &\text{ otherwise } \\
        \end{cases}
        $$

        and for $j\leq v-1$, $\delta$ goes from the end of $Y_j$ to $X_{j+1}$
            i.e. for any $1\leq i \leq v'$
        $$
            \delta(q^{Y_j}_i,x,z_0) =
        \begin{cases}
        (q^{X_{j+1}}_1,1z_0) &\text{ if } x=1\\
        (q^{X_{j+1}}_0,0z_0) &\text{ if } x=0 .
        \end{cases}
        $$
    and at the end of $Y_v$, a new $A$ zone starts, thus for any $1\leq i \leq v'$
    $$
            \delta(q_i^{Y_v},x,z_0) =
        \begin{cases}
        (q^a_1,z_0) &\text{ if } x=1\\
        (q^a_0,z_0) &\text{ if } x=0 .
        \end{cases}
        $$

        Once in the $q_e$ state, $\delta$ never leaves it, i.e.
        $$
            \delta(q_e,x,y) = (q_e,y)
        $$

        The output  function outputs the input on every states, except on states $q^{Y_j}_1, \ldots, q^{Y_j}_{v'} \quad$            $(j=1,\ldots,v)$
        where for $1\leq i <v'$
        $$
            \nu(q^{Y_j}_{i},b,y) = \lambda
        $$
        and
        $$
            \nu(q^{Y_j}_{v'},b,y) = 0.
        $$
       \begin{flushright}
    $\Box $
\end{flushright}


\end{document}